\documentclass[
  aps,
  prl,
  reprint,
  superscriptaddress,
  amsmath,
  amssymb,
  floatfix,
  nobibnotes,
  nofootinbib
]{revtex4-2}
 
\usepackage[T1]{fontenc}
\usepackage[dvipsnames]{xcolor}
\usepackage{bm}
\usepackage{mathrsfs}
\usepackage{cancel}
\usepackage{extarrows}
\usepackage{esint}
\usepackage{amsthm}
\usepackage{graphicx}
\usepackage{booktabs}
\usepackage[most]{tcolorbox}
\usepackage{physics}
\usepackage[hidelinks]{hyperref}
\usepackage{orcidlink}
\usepackage[normalem]{ulem}

\newcommand{\be}{\begin{equation}}
\newcommand{\ee}{\end{equation}}

\def\bea{\begin{eqnarray}}
\def\eea{\end{eqnarray}}

\begin{document}

\title{An Electromagnetic Ergoregion Limits the Reconnection-Driven Penrose Process}

\author{Qian Wan}
\affiliation{School of Physics and Optoelectronics, South China University of Technology, Guangzhou 510641, P. R. China}

\author{Yehui Hou}
\email{contact: yehuihou@sjtu.edu.cn}
\affiliation{Tsung-Dao Lee Institute, Shanghai Jiao-Tong University, Shanghai, 201210, P. R. China}

\author{Yu Song}
\affiliation{Shanghai Astronomical Observatory, Chinese Academy of Sciences, Shanghai, 200030, P. R. China}

\author{Jing-Ze Xia}\affiliation{Tsung-Dao Lee Institute, Shanghai Jiao-Tong University, Shanghai, 201210, P. R. China}

\author{Yosuke Mizuno}
\affiliation{Tsung-Dao Lee Institute, Shanghai Jiao-Tong University, Shanghai, 201210, P. R. China}
\affiliation{School of Physics \& Astronomy, Shanghai Jiao-Tong University, Shanghai, 200240, P. R. China}
\affiliation{Key Laboratory for Particle Physics, Astrophysics and Cosmology (MOE), Shanghai Key Laboratory for Particle Physics and Cosmology, Shanghai Jiao-Tong University, Shanghai, 200240, P. R. China}

\author{Bin Chen}
\email{contact: chenbin1@nbu.edu.cn}
\affiliation{Institute of Fundamental Physics and Quantum Technology, \& School of Physical Science and Technology, Ningbo University, Ningbo, Zhejiang, 315211, China}
\affiliation{Zhejiang Key Laboratory of Extreme Universe, \& BINGO Center, 
    Ningbo University, Ningbo, Zhejiang 315211, China }
\affiliation{Center for High Energy Physics, \& School of Physics, Peking University, No.5 Yiheyuan Rd, Beijing, 100871, P. R. China}

\begin{abstract}
Magnetic reconnection taps black-hole rotational energy via negative-energy ejecta, but field geometry itself can forbid such ejection inside the spacetime ergoregion.
We identify a covariant electromagnetic (EM) ergoregion for general degenerate, magnetically dominated fields in stationary spacetimes.
Outside the EM ergoregion, even ultrarelativistic field-guided ejecta cannot attain negative Killing energy. For Alfv\'enic reconnection outflows, the magnetization threshold scales inversely with the transverse distance to the EM ergosurface.
The mass-loaded Kerr model places the EM ergosurface at about 75\% of the radial interval from the horizon to the spacetime ergosurface, close to the time-averaged boundary inferred from a dynamical simulation.
Together, these results show that field geometry determines where negative-energy ejection is possible, while magnetization sets the threshold for realizing it.
\end{abstract}

\maketitle

\noindent\textit{Introduction.—}
The rotational energy of a black hole can be extracted through negative-energy matter, EM torques, or wave superradiance \cite{Penrose:1971uk,Starobinskii:1973vzb, Blandford:1977ds,Komissarov:2005wj, Komissarov:2008yh, Lasota:2013kia,Brito:2015oca, Noda:2019mzd}, powering a variety of high-energy astrophysical phenomena. 
Magnetic reconnection provides a plasma-mediated realization of the Penrose process:
oppositely directed exhausts can produce a negative-energy plasmoid captured by the black hole, accompanied by a positive-energy counterpart that can carry the released energy
outward \cite{Koide:2008xr,Comisso:2020ykg}. 
General-relativistic descriptions of reconnection, together with studies of upstream
magnetic geometry and multidimensional ejection, have further clarified the local conditions for this mechanism \cite{Asenjo:2015xva, Asenjo:2017gsv, Chen:2024ggq, Shen:2024plw, Camilloni:2024tny}.

\begin{figure}[htbp]
    \centering
    \hspace*{-1mm}
    \includegraphics[width=3.4in,trim={2mm 3.5mm 1mm 1mm},clip]{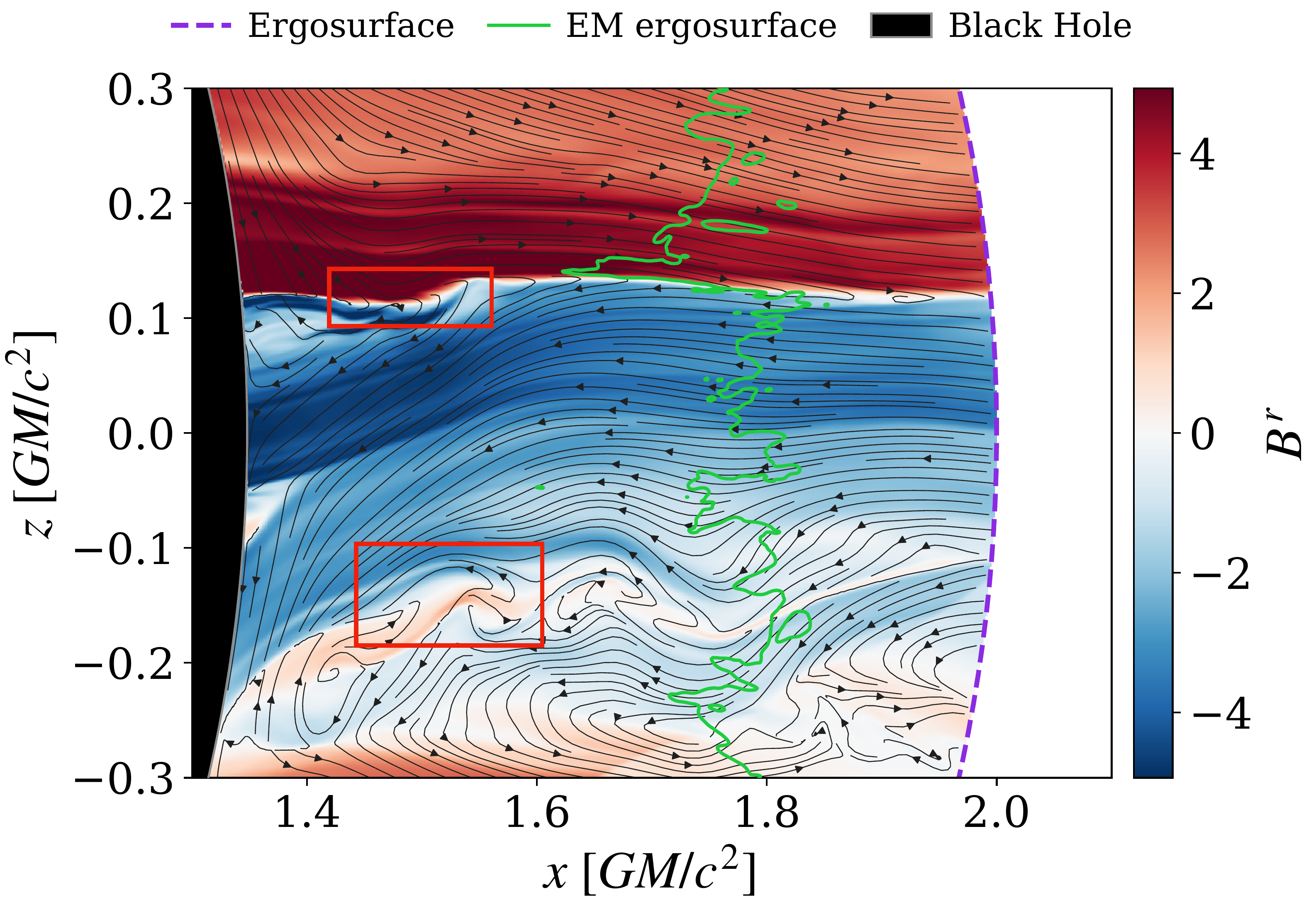}
\caption{Magnetic-field configuration near a Kerr black hole with spin
$a=0.9375$, obtained from a two-dimensional GRMHD simulation on a
$4800\times2400$ grid in Kerr--Schild coordinates. Arrows indicate the
magnetic-field direction, while the color map shows its radial component.
The dashed purple and solid green curves mark the spacetime ergosurface
and the time-averaged EM-ergoregion boundary, respectively. Red boxes highlight candidate reconnection structures.}
    \label{fig_kharma}
\end{figure}

In a dynamical magnetosphere, reconnection occurs throughout the system—in near-equatorial current sheets and jet sheaths, and along disk--magnetosphere separatrices \cite{Ripperda:2020bpz,ElMellah:2021tjo,Bransgrove:2021heo,Blandford:2022ewb, ElMellah:2023sun,Sironi:2025kgn}. Candidate reconnecting layers therefore appear at many latitudes in an accreting magnetosphere (Fig.~\ref{fig_kharma}).
Yet reconnection inside the spacetime ergoregion does not by itself ensure rotational-energy extraction:
the same upstream magnetic field that supports energy release also guides the exhausts, restricting their directions of motion; even an arbitrarily energetic ejection may therefore fail to reach the negative-Killing-energy state required by the Penrose process. 
The basic question is thus not only whether reconnection can accelerate plasma, but where the EM-field geometry permits negative-energy generation in the first place.
Astrophysically, this location therefore sets how much rotational energy can be loaded into the escaping plasmoids, and hence the energetics of hot spots and multiwavelength flares \cite{Ripperda:2020bpz,ElMellah:2023sun,Aimar:2023kzj,Galishnikova:2022mjg}.

In this Letter, we show that negative-energy accessibility is governed by a geometric criterion that depends on the field configuration, but not on the field strength or plasma details.
We begin with a degenerate, magnetically dominated EM field of arbitrary geometry in a stationary spacetime, where the field defines timelike two-dimensional field sheets, within which the plasma four-velocity and the comoving magnetic field lie \cite{Gralla:2014yja}. 
We establish a local criterion for negative-energy field-guided motion 
and show that, for a positive-energy plasma flow, negative-energy motion becomes possible, in the ultrarelativistic limit, precisely when the projection of the stationary Killing vector onto the sheet is spacelike.

The resulting \emph{EM ergoregion} lies within the spacetime ergoregion but need not fill it. Outside it the reconnection-driven Penrose process is forbidden, no matter how energetic the outflow is. Since the condition is invariant under a rescaling of the field, a stronger field cannot enlarge this region; only the field geometry can. 
In the stationary, axisymmetric case, the EM ergosurface is determined explicitly by the field-line angular velocity, the relative poloidal-field strength, and the poloidal current.
These quantities can be evaluated directly from GRMHD data.
For a mass-loaded magnetosphere with a split-monopole poloidal field in the Kerr spacetime, this region spans approximately $3/4$ of the radial interval from the horizon to the spacetime ergosurface.

For this magnetosphere, we determine the local energy-extraction efficiency and the energy threshold for negative-energy production.
The upstream state is obtained from a cold, ideal-GRMHD inflow selected by regularity at the fast-magnetosonic point \cite{Nitta:1991ui,beskin1992}.
A localized fast-reconnection layer then relates the outflow speed to the upstream magnetization. 
As the reconnection site approaches the EM ergosurface from within, the magnetization required for negative-energy production diverges. 
We further examine whether the accompanying positive-energy counterparts escape in the ballistic approximation.
The accessibility of negative-energy states is thus geometric, whereas their realization is dynamical.
Throughout, we use metric signature $(-,+,+,+)$ and units with $G=c=1$.

\medskip
\noindent\textit{General negative-energy accessibility.—} We consider a four-dimensional, stationary,   asymptotically flat spacetime with stationary Killing vector \(\xi^\mu\), normalized at infinity. Let \(F_{\mu\nu}\) be a smooth, closed electromagnetic two-form satisfying
\begin{equation}
F_{\mu\nu}{}^\star F^{\mu\nu}=0\,,
\quad
F^2\equiv F_{\mu\nu}F^{\mu\nu}>0\,.
\end{equation}
These conditions ensure that the kernel of \(F_{\mu\nu}\), \(\ker F\equiv\{X^\mu\mid F_{\mu\nu}X^\nu=0\}\), is a two-dimensional timelike plane \cite{rainich1925electrodynamics,misner1957classical}. Choose a future-directed unit timelike vector \(U^\mu\in\ker F\), representing the ideal-plasma frame, and define \(b^\mu={}^\star F^{\nu\mu}U_\nu\), with \(b^2=F^2/2\). The unit vector \(\ell^\mu=b^\mu/\sqrt{b^2}\), together with \(U^\mu\), spans \(\ker F\). The closure condition \(dF=0\) makes this distribution integrable, with its integral surfaces defining the field sheets \cite{Gralla:2014yja, Gralla:2018kif}.

We consider a locally symmetric ejection consisting of two equal-rest-mass ejecta moving in opposite directions along the field sheet, with equal speed \(v_{\rm out}\) measured in the \(U^\mu\) frame. Their four-velocities are 
\begin{equation}\label{eq:pairspeed}
u^\mu_{(s)}=\gamma_{\rm out}
\left(U^\mu+s v_{\rm out}\ell^\mu\right)\,,
\quad s=\pm1\,,
\end{equation}
where \(\gamma_{\rm out}=(1-v_{\rm out}^2)^{-1/2}\) is the Lorentz factor. The corresponding specific Killing energies are \(\mathcal E_s=-\xi_\mu u^\mu_{(s)}\).
For a future-directed flow with positive
Killing energy, \(-\xi\cdot U>0\), the pair has positive total bulk energy, \(\mathcal E_++\mathcal E_-=-2\gamma_{\rm out}(\xi\cdot U)>0\), and one of the ejecta can acquire negative energy if and only if \(v_{\rm out}^2(\xi\cdot\ell)^2>(\xi\cdot U)^2\). 

To express the negative-energy condition in terms of the spacetime and EM-field geometries, we introduce
\begin{equation}
\mathcal Q_F\,\equiv\, \xi^2-
\frac{(\xi^\mu F_{\mu\alpha})
(\xi^\nu F_\nu{}^\alpha)}{F^2/2}\,,
\end{equation}
The subtracted term is the squared norm of the component of \(\xi^\mu\) normal to the field sheet, so \(\mathcal Q_F\) is the squared norm of the projection of \(\xi^\mu\) onto \(\ker F\), i.e., \(\mathcal Q_F=(\xi\cdot\ell)^2-(\xi\cdot U)^2\). Hence, the negative-energy condition becomes
\begin{equation}\label{eq:finbound}
\mathcal Q_F\,>\,
\frac{(\xi\cdot U)^2}{\gamma_{\rm out}^2-1}\,.
\end{equation}
The left-hand side depends only on the spacetime and the field geometry, whereas the right-hand side depends on the outflow only through its Lorentz factor.
In the ultrarelativistic limit with $\gamma_{\rm out}\to\infty$, the ejecta approach two null directions on the field sheet, and Eq.~\eqref{eq:finbound} reduces to \(\mathcal Q_F>0\).
Hence, for a fixed spacetime and field geometry, the region $\mathcal Q_F>0$ does not depend on the plasma state or on the outflow speed.
We thus define 
\bea
\mathscr{E}_{\rm EM}\,\equiv\, \{\,x \,\mid \mathcal Q_F(x)>0\,\}
\eea
as the EM ergoregion\footnote{The phrase ``electromagnetic ergoregion'' has been used before in a Kaluza--Klein discussion of staticity conditions for nonrotating black holes \cite{Carter:1988em}; here it denotes the region $\mathcal Q_F>0$.}, with \(\mathcal Q_F=0\)
defining its boundary, \(\mathscr S_{\rm EM}\equiv\partial\mathscr E_{\rm EM}\). This region contains every negative-energy outcome of the pair ejection described by Eq.~\eqref{eq:pairspeed}. At finite outflow speed, Eq.~\eqref{eq:finbound} imposes an additional Lorentz-factor threshold within $\mathscr{E}_{\rm EM}$.

Since the orthogonal complement of the timelike field sheet is spacelike, \(\mathcal Q_F\leq\xi^2\), so $\mathscr{E}_{\rm EM}$ is contained within the spacetime ergoregion, \(\xi^2>0\). 
The projection of the Killing vector onto a field sheet can remain timelike even inside the spacetime ergoregion, precluding negative-energy field-aligned motion. 
This criterion is local and covariant, independent of the overall field normalization, and requires no EM-field symmetries.

This criterion complements the energy limits on the particle Penrose process \cite{Wald:1974kya}: rather than constraining the energy gain in a breakup, it determines whether
field-guided ejecta can access negative Killing energy in the first place. 
The Blandford--Znajek process has been formulated in terms of field sheets with an effective ergoregion \cite{okamoto1992evolution, Kinoshita:2017mio}; Alfvén-wave superradiance is instead governed by the light surface and the wave frequencies \cite{Noda:2019mzd, Noda:2025ota}. 
$\mathscr{E}_{\rm EM}$ is distinct from these descriptions:
it marks the local accessibility of negative-energy particle motion, rather than characterizing the energy content of the EM field itself or GRMHD perturbations. 

\medskip
\noindent\textit{Dynamical realization by magnetic reconnection.—}
We now ask whether reconnection can supply the outflow speed required by Eq.~\eqref{eq:finbound}.
We consider a localized nonideal layer aligned with the upstream magnetic field, across which the reconnecting field reverses direction \cite{Priest1986,Titov1997}.
Assuming approximately symmetric upstream conditions, magnetic-energy release drives oppositely directed exhausts along the upstream field direction in the comoving frame \cite{1964NASSP..50..425P,Lyubarsky:2005zt}.
The layer is treated as a pointlike X-point on magnetospheric scales. Discrete plasmoids may form through tearing-mode or plasmoid instabilities.

For relativistic reconnection, the outflow is typically Alfvénic. In terms of the upstream magnetization, \(\sigma=b^2/\rho\) with $\rho$ the rest-mass density, the Alfvén speed satisfies \(v_A^2=\sigma/(1+\sigma)\). Thus, for relativistic ejections, \(\gamma_{\rm out}^2-1\sim\sigma\), up to model-dependent factors. Combining this relation with Eq.~\eqref{eq:finbound} yields the approximate magnetization requirement \(\sigma\gtrsim(\xi\cdot U)^2/\mathcal Q_F\).
The required magnetization diverges as the reconnection site approaches $\mathscr{S}_{\rm EM}$  from within, \(\mathcal Q_F\rightarrow0^+\). 

Near a regular point of \(\mathscr S_{\rm EM}\), where the value \(q_F\equiv\partial_\perp\mathcal Q_F|_{\mathcal Q_F = 0}\neq0\),
we can expand $\mathcal Q_F = q_F\, d_{\perp} + \mathcal{O}(d_{\perp}^2)$, with $d_{\perp}$ the signed transverse distance. The energy threshold required to produce negative energy is then described by
\bea
\gamma_{\rm out}^2 \,\approx\, \frac{(\xi\cdot U)_{\rm em}^2}{q_F \,d_{\perp}} \,,
\eea
with the subscript ``em'' denoting the value exactly on $\mathscr{S}_{\rm EM}$. Combined with a typical relation in relativistic reconnection \cite{Lyubarsky:2005zt, Sironi:2014jfa}, $\gamma_{\rm out} \propto \sqrt\sigma$, we obtain $\sigma\propto d_\perp^{-1}$. The exponent follows from the linear vanishing of $\mathcal Q_F$ at the boundary, while the coefficient is fixed by the local field geometry and reconnection model. Close to $\mathscr{S}_{\rm EM}$ the required magnetization is arbitrarily large, so the process is suppressed at any finite $\sigma$.
For $\mathcal Q_F\leq0$, negative-energy ejection is forbidden regardless of the outflow speed.

\medskip
\noindent\textit{Stationary, axisymmetric magnetosphere.—}
To evaluate the energy-extraction threshold in black-hole magnetospheric models, we specialize to a stationary, axisymmetric EM field in a circular spacetime, with line element $\dd s^2=g_{tt}\dd t^2+2g_{t\phi}\dd t\dd\phi +g_{\phi\phi}\dd\phi^2+g_{rr}\dd r^2+g_{\theta\theta}\dd\theta^2$,
where the metric components depend only on $r$ and $\theta$. This class includes the Kerr family in Boyer--Lindquist coordinates \cite{kerr1965new,Carter:1968rr}.
Magnetospheric solutions determine the upstream field and flow that set the plasmoid initial conditions \cite{Gammie:2003rj,McKinney:2012vh,EventHorizonTelescope:2019pcy,Dhruv:2024igk}.

For a stationary, axisymmetric, degenerate EM field, the Faraday tensor can be written in terms of three scalar functions:
$F= \kappa^{-1}\sqrt{-g} \,J\,dr\wedge d\theta
+d\psi\wedge\left[d\phi-\Omega_F\,dt\right]$,
with \(\kappa\equiv g_{t\phi}^2-g_{tt}g_{\phi\phi}\). 
The flux function \(\psi\equiv A_\phi\) labels the magnetic surfaces and the associated family of field sheets; the field-line angular velocity \(\Omega_F=\Omega_F(\psi)\) fixes the toroidal rotation \cite{thorne1982electrodynamics,Gralla:2014yja}, and the helical vector \(\chi_F^\mu=(\partial_t)^\mu+ \Omega_F(\partial_\phi)^\mu\) is tangent to the field sheet;
\(J\) encodes the poloidal current, which generally varies along a field line, becoming $J = J(\psi)$ only in the force-free limit \cite{Gralla:2014yja}.

Under the symmetries of the magnetosphere, the EM ergoregion is characterized by a simple relation
\bea
\label{eq:magnetosphere_bound}
\begin{gathered}
g_{tt}\,b^2 > \kappa\,\Omega_F^2 B_{\rm p}^2\,,\\[3pt]
b^2 = k_0 B_{\rm p}^2
+ J^2/\kappa\,,\,\,
B_{\rm p}^2 =\frac{g^{rr}(\partial_r\psi)^2
+g^{\theta\theta}(\partial_\theta\psi)^2
}{\kappa}\,,
\end{gathered}
\eea
where $k_0=-\chi_F^2 $; \(B_{\rm p}\) denotes the poloidal-field strength. 
The location of $\mathscr{S}_{\rm EM}$ is determined by the field-line rotation, the relative poloidal-field strength
\((B_{\rm p}^2/b^2)\) and the poloidal current. 
The criterion is independent of the overall EM-field normalization and provides a local geometric bound on negative-energy field-guided motion.
 Because $\Omega_F$ and $J$ are constrained by horizon regularity and the global flow structure, the magnetized plasma configuration plays a central role in shaping the EM ergoregion
\cite{meier2001magnetohydrodynamic, Narayan:2003by,
Tchekhovskoy:2011zx, Penna:2013rga, Mizuno:2022vqa}. For a purely toroidal field, or zero-rotation field ($\Omega_F=0$), $\mathscr{E}_{\rm EM}$ reduces to the spacetime ergoregion.

\medskip
\noindent\textit{A Kerr magnetosphere realization.—}
We illustrate the above formulation in a Kerr spacetime
with black-hole mass $M$ and dimensionless spin $a\in(0,1)$.
The near-horizon EM field can be well described by a simplified model with a split-monopole poloidal field \cite{Komissarov:2004qu,Komissarov:2005wj,Tchekhovskoy:2009da,
Parfrey:2018dnc,Ripperda:2021zpn},
whose field sheets are labeled by the polar angle $\theta_+\in(0,\pi)$.
The flux functions are given by
$\psi(\theta_+)=\psi_0(1-\abs{\cos\theta_+})$, and
$\Omega_F(\theta_+)=a(2r_++r_+^2/M+Ma^2\cos^2\theta_+)^{-1}$, where $r_+$ is the event-horizon radius
\footnote{We estimate $\Omega_F$ by matching the Znajek
horizon current to the asymptotic Michel-type current,
$J_H=J_\infty=-\Omega_F\psi_0\sin^2\theta_+$
\cite{Znajekmnras,Michel1969,Pu:2015}.
This force-free prescription is used only to fix $\Omega_F$ and is not imposed as a global current constraint on the ideal-GRMHD solution; the normalization $\psi_0$ drops out of all results.}. Departures from this prescribed radial geometry
due to trans-field force balance are minor at the horizon scale
\cite{Huang:2019wqv}.

The upstream state is obtained from a stationary, axisymmetric, cold, ideal GRMHD inflow.
Plasma is loaded at \(r=r_{\rm load}\), where the corotation potential \(k_0\) has a local maximum
along the streamline \cite{Song:2025mhj}.
The flow dynamics is constrained by the Bernoulli equation with conserved mass loading, total specific energy and angular momentum \cite{BekensteinNew, camenzind1986}.
The loading and horizon boundary conditions, together
with fast-magnetosonic regularity, select a continuous, positive-density, one-parameter family of inflow solutions 
\cite{Takahashi1990,Takahashi1991}.
We parameterize this family by the horizon magnetization $\sigma_H$.
At each reconnection radius, the local upstream magnetization $\sigma$ sets the ejection speed through the fast-reconnection prescription $v_{\rm out}=[\sigma/(\Lambda+\sigma)]^{1/2}$ with $\Lambda=3.5$ \cite{Liu:2016sqd,Chen:2024ggq}. The local upstream velocity, field, and density, together with this ejection speed, then fix the energies of the ejected plasmoids as functions of the reconnection position.

Among the configurations considered, moving the reconnection point toward the horizon increases the larger energy $\varepsilon_{\rm high}$ and decreases the smaller one
$\varepsilon_{\rm low}$, with the latter becoming negative
once a threshold within $\mathscr{E}_{\rm EM}$ is crossed.
The extraction efficiency $\eta_{\rm ext}\equiv\varepsilon_{\rm high}/
(\varepsilon_{\rm high}+\varepsilon_{\rm low})$ increases toward a horizon limit.
Higher magnetization enhances both $\varepsilon_{\rm high}$
and $\eta_{\rm ext}$, though the efficiency depends
only weakly on magnetization for $\sigma_H\gtrsim10^2$,
approaching its high-magnetization limit.
Moving toward the equatorial plane increases $\eta_{\rm ext}$, with stronger
angular dependence at higher spin.
For $\sigma_H=100$, the maximum efficiency at the horizon is
$1.1$ for $a=0.9$, $\theta_+=\pi/3$, and $1.35$
for $a=0.99$, $\theta_+=\pi/2$.
These values remain below the limiting value of
$\eta_{\rm ext}=1.5$ obtained in the extremal Kerr limit in equatorial ballistic models \cite{Comisso:2020ykg,Chen:2024ggq}.

\begin{figure}[htbp]
    \centering
    \hspace*{-1cm}
    \includegraphics[width=3.2in,trim={2mm 5mm 1mm 1mm},
        clip]{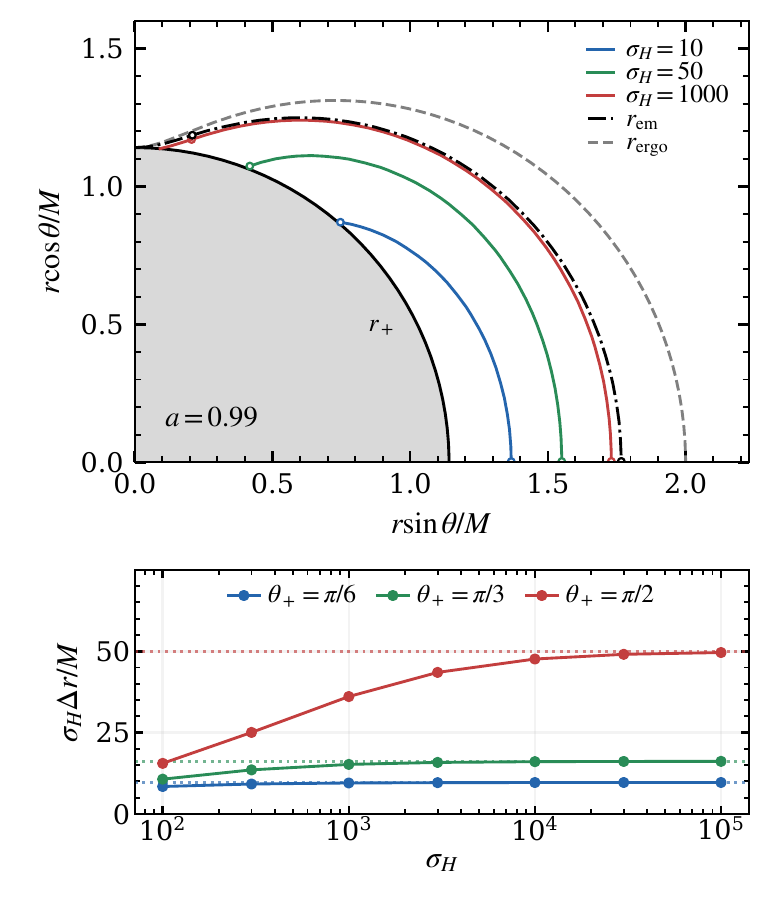}
    \caption{\textbf{Top:} Spatial thresholds for negative-energy production
for different magnetizations $\sigma_H$ (colored curves)
and $\mathscr{S}_{\rm EM}$ (dash-dotted line) in the northern
meridional plane. Each boundary is constructed from trans-fast
inflow solutions at fixed $\sigma_H$ and varying colatitude.
\textbf{Bottom:} Rescaled radial gap $\sigma_H\Delta r$ between $\mathscr{S}_{\rm EM}$
and the threshold for negative-energy production as a function of $\sigma_H$
at selected colatitudes. The dotted line indicates
$\Delta r\propto\sigma_H^{-1}$.
}
    \label{fig_rem}
\end{figure}

\medskip
\noindent\textit{Critical radii.—}
The EM ergosurface $\mathscr{S}_{\rm EM}$ sets a universal geometric limit, while finite magnetization further restricts where negative-energy ejecta can be produced.
Fig.~\ref{fig_rem} shows the production boundaries defined by equality in Eq.~\eqref{eq:finbound} for the adopted mass-loaded Kerr magnetosphere. For $\sigma_H=10$ and $50$, the boundaries terminate at $\theta_+=40.3^\circ$ and $20.9^\circ$, respectively, leaving a polar range in which negative-energy production is inaccessible. 
Increasing the magnetization therefore enlarges both the angular coverage and the radial extent
of the production region.
At high magnetization, the production boundary approaches $\mathscr{S}_{\rm EM}$. Their radial separation $\Delta r $ obeys
\begin{equation}
\Delta r 
\,\propto \, M\sigma_H^{-1}\,,
\quad \sigma_H\gg1\,.
\end{equation}
The lower panel of Fig.~\ref{fig_rem} resolves this approach by plotting the rescaled gap:
its asymptotic plateaus identify the inverse-magnetization scaling. The onset of this regime depends on colatitude and occurs at lower $\sigma_H$ closer to the rotation axis.
For example, at $\theta_+=30^\circ$, the scaling is already established for $\sigma_H\gtrsim10^3$.

To locate $\mathscr{S}_{\rm EM}$, denoted by $r = r_{\rm em}(\theta)$, we solve
the equation given by Eq.~\eqref{eq:magnetosphere_bound} for its outer radial root.
For the adopted split-monopole prescription, the equation
is quintic in $r$ and reduces to a cubic on the equatorial
plane. A spin expansion at fixed $\theta_+\in(0,\pi)$ gives
\begin{equation}\label{eq:fext}
\begin{aligned}
\frac{r_{\rm em}-r_+}{r_{\rm ergo}-r_+}
&={}1 - q_\Omega^2 + \frac{3}{4}a^2q_\Omega^2(1-q_\Omega)^2\sin^2\theta_+
+\mathcal{O}(a^4)\,,
\end{aligned}
\end{equation}
where $q_\Omega = \Omega_F / \Omega_H$. For horizon-threading field lines, $q_\Omega \simeq 0.5$ via the horizon regularity condition \cite{Znajekmnras}. Thus, $\mathscr{E}_{\rm EM}$ spans $\simeq 3/4$ of the radial interval between the horizon and the spacetime ergosurface, and this fraction persists even at high spin (Table~\ref{tab1}).
Because the boundary is geometric, raising the magnetization can populate the accessible region but cannot extend it.

\begin{table}
    \centering
    \begingroup
    \renewcommand{\arraystretch}{1} 
    \setlength{\tabcolsep}{5pt}    
    \setlength{\aboverulesep}{2pt}    
    \setlength{\belowrulesep}{2pt}     
    \begin{tabular*}{\linewidth}
        {@{\extracolsep{\fill}}ccccccc@{}}
        \toprule
        $\theta_+$
        & \multicolumn{3}{c}{$\pi/3$}
        & \multicolumn{3}{c}{$\pi/2$} \\
        $a$ & 0.8 & 0.9 & 0.99 & 0.8 & 0.9 & 0.99 \\
        \midrule
        $\%$
        & 75.83 & 75.27 & 75.42 & 75.00 & 73.41 & 73.22 \\
        \bottomrule
    \end{tabular*}
    \endgroup
    \caption{Numerically evaluated radial extent of $\mathscr{E}_{\rm EM}$, expressed as $(r_{\rm em}-r_+)/(r_{\rm ergo}-r_+)$, for the Kerr magnetosphere with different black-hole spins and polar angles of the reconnection point.}
    \label{tab1}
\end{table}

The time-averaged contour inferred from the simulation in Fig.~\ref{fig_kharma} lies near the EM ergosurface predicted by Eq.~\eqref{eq:fext}. This comparison illustrates the applicability of 
\(\mathcal Q_F\) as a geometric diagnostic of negative-energy accessibility in dynamical accretion flows.
The instantaneous EM ergosurface, $b^2g_{tt}=F_{t\alpha}F_t{}^\alpha$, develops a complex structure, with loops forming and disappearing. 
To isolate the main structure, we consider the EM boundary inferred from temporally averaged field quantities over $500M$.

\begin{figure}[htbp]
	\centering
	\hspace*{-0.7cm}
	\includegraphics[width=3.2in,trim={2mm 5mm 1mm 1mm},
	clip]{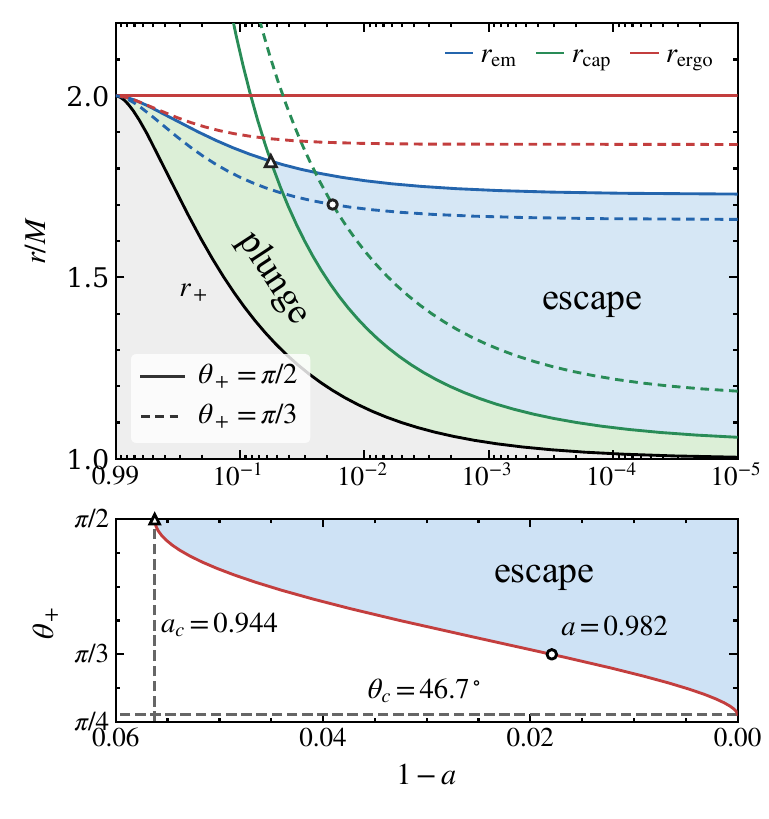}
\caption{\textbf{Top:} Limiting radii for negative-energy production and ballistic escape as functions of black-hole spin at $\theta_+=\pi/3$ and $\pi/2$.
For $\theta_+=\pi/2$, the blue and green regions allow
negative-energy production with and without escape of the
positive-energy counterpart, respectively.
\textbf{Bottom:} Parameter region allowing both negative-energy
production and positive-energy escape in the high-magnetization limit.}
	\label{fig_crit}
\end{figure}

The positive-energy ejecta can carry the extracted energy to infinity or deposit it as heat in the surrounding plasma. 
We focus here on direct, ballistic escape to infinity as a criterion for completing the Penrose process \cite{Chen:2024ggq, YuChih:2026ios}, in which plasmoids follow geodesics specified by their
formation sites and initial four-momenta.
For the ultrarelativistic ejecta, unstable spherical photon orbits
define the capture--escape boundary.
Matching the initial ejection direction to the critical null direction determines the escape region, $r_{\rm ergo} > r > r_{\rm cap}$.
Fig.~\ref{fig_crit} shows the critical radii, $r_{\rm em}$, $r_{\rm cap}$, and the
overlap between the regions allowing negative-energy production and positive-energy escape (shaded blue). 
Both radii decrease with increasing spin. 
As the reconnection site approaches the equatorial plane, however, $r_{\rm em}$ increases whereas $r_{\rm cap}$ decreases. 
The overlap thus widens toward the equator and exists only above a colatitude-dependent minimum spin.

The lower panel in Fig.~\ref{fig_crit} shows the polar interval allowing the complete Penrose process, which depends only on the spin. 
As the spin increases from a threshold $a_c\simeq0.944$, the polar interval expands from a narrow range around the equatorial plane to a maximal interval $\theta \in (\theta_c, 180^\circ-\theta_c)$, with $\theta_c\simeq 46.7^\circ$, in the extremal Kerr limit \footnote{This differs from the value $a\simeq0.8$ reported in Ref.~\cite{YuChih:2026ios}, where the ejection direction is prescribed by hand and the magnetospheric constraint is not considered.}. The escape condition thus selects a subset of the EM ergoregion that supports energy extraction to infinity. 

\medskip
\noindent\textit{Summary and Outlook.—}
The spacetime ergoregion determines where negative energy is allowed by gravity, 
whereas the EM field determines where negative-energy motion is accessible to the plasma.
The EM field that powers reconnection also guides
the ejecta, restricting their access to black-hole
rotational energy. We have identified this restriction
through the EM ergoregion, which separates the geometric
accessibility of negative bulk Killing energy from its
dynamical realization. At fixed field-sheet geometry,
faster ejection cannot overcome geometric exclusion;
inside the region, an energetic outflow is still
required. In the adopted mass-loaded Kerr model, this
distinction becomes quantitative: the production boundary
approaches the EM ergosurface with a radial gap
$\Delta r\propto M\sigma_H^{-1}$ at high magnetization.
Escape of the positive-energy counterpart
further adds a separate, global constraint.

Independent of a reconnection prescription and requiring
neither a stationary nor an axisymmetric EM field, the local criterion $\mathcal Q_F$ provides a
geometric reference for studying reconnection in
time-dependent, three-dimensional magnetospheres,
including disk--jet interfaces
\cite{Fendt:2005fq,Dihingia:2021ncv,Nathanail:2021jbn,Jiang:2023ygs}.
Comparing it with reconnection sites and ejecta trajectories
could reveal when and how often the allowed negative-energy
states are populated, while testing the field-guided
approximation near nonideal regions.
Combined with horizon energy-flux diagnostics, these measurements could help
distinguish local plasma energization from events that
contribute to black-hole rotational-energy extraction.
Since the $3/4$ fraction depends weakly on spin (Table~\ref{tab1}), it provides a simple
reference value against which reconnection-powered flare models can be compared.

Connecting such extraction events to flares or jet
variability requires following energy transport beyond
the ballistic approximation, including magnetic forces,
radiative losses, and interactions with the surrounding
plasma \cite{Xia:2026nkt,Li:2026yhd}.
The remaining challenge is to determine how much of the
extracted energy reaches large distances, and in what
observable form.

\medskip
\noindent\textit{Acknowledgments.—}
We thank Ye Shen and Zhenyu Zhang for helpful discussions.
The work is partly supported by NSFC Grant Nos. 12588101, 12275004, 11735001,  12547123, 12273022, 1260050079 and 12511540053. YM is also supported by the National Key R\&D Program of China No. 2023YFE0101200 and the Shanghai Municipality Orientation Program of Basic Research for International Scientists No. 22JC1410600. 

\medskip

\noindent \textit{Statement on the use of AI.—}
The ideas, methodology, and manuscript were developed by the authors. OpenAI's GPT-6 was used to assist with language editing and consistency checks. The authors take full responsibility for the content of the manuscript.

\medskip
\twocolumngrid
\bibliographystyle{utphys}
\bibliography{refs}

\newpage
\onecolumngrid
\begin{center}
    \textbf{\large Supplemental material: escape boundaries for positive-energy plasmoids}
\end{center}
The escape condition for positive-energy plasmoids is determined in the ballistic approximation. After leaving the reconnection point, each plasmoid is assumed to experience negligible nongravitational forces, so that its center of mass follows a geodesic with initial data specified by the reconnection model. In the ultrarelativistic, high-magnetization limit, this trajectory is approximated by a null geodesic launched along $N_{\pm}^\mu \equiv U^\mu\pm\ell^\mu$. This approximation neglects subsequent magnetic forces; high magnetization alone does not imply dynamical decoupling. The conserved axial angular momentum and Carter constant, normalized by the Killing energy and its square, respectively, are
\begin{equation}
	\lambda_{\pm} =-\frac{(N_{\pm})_{\phi}}{(N_{\pm})_{t}}\,,\quad
	q_{\pm}=\frac{(N_{\pm})_{\theta}^2}{(N_{\pm})_{t}^2}+\cos^2\theta_+
	\left(\frac{\lambda_\pm^2}{\sin^2\theta_+}-a^2\right)\,,
\end{equation}
where $\pm$ labels the ejection direction and all quantities on the right-hand side are evaluated at launch. Hereafter, we set $M=1$. The radial motion is governed by the quartic potential
$R(r)=[r^2+a^2-a\lambda_\pm]^2-\Delta[(\lambda_\pm-a)^2+q_\pm]$, where $\Delta = r^2-2r+a^2$.
Its roots, together with the initial radial direction, determine whether the plasmoid escapes or is captured. In particular, an initially inward-moving plasmoid can escape if it encounters an exterior radial turning point in an allowed region connected to infinity.

\begin{figure}[htbp]
	\centering
	\includegraphics[width=6.5in,trim={5mm 8.2mm 1mm 1mm},clip]{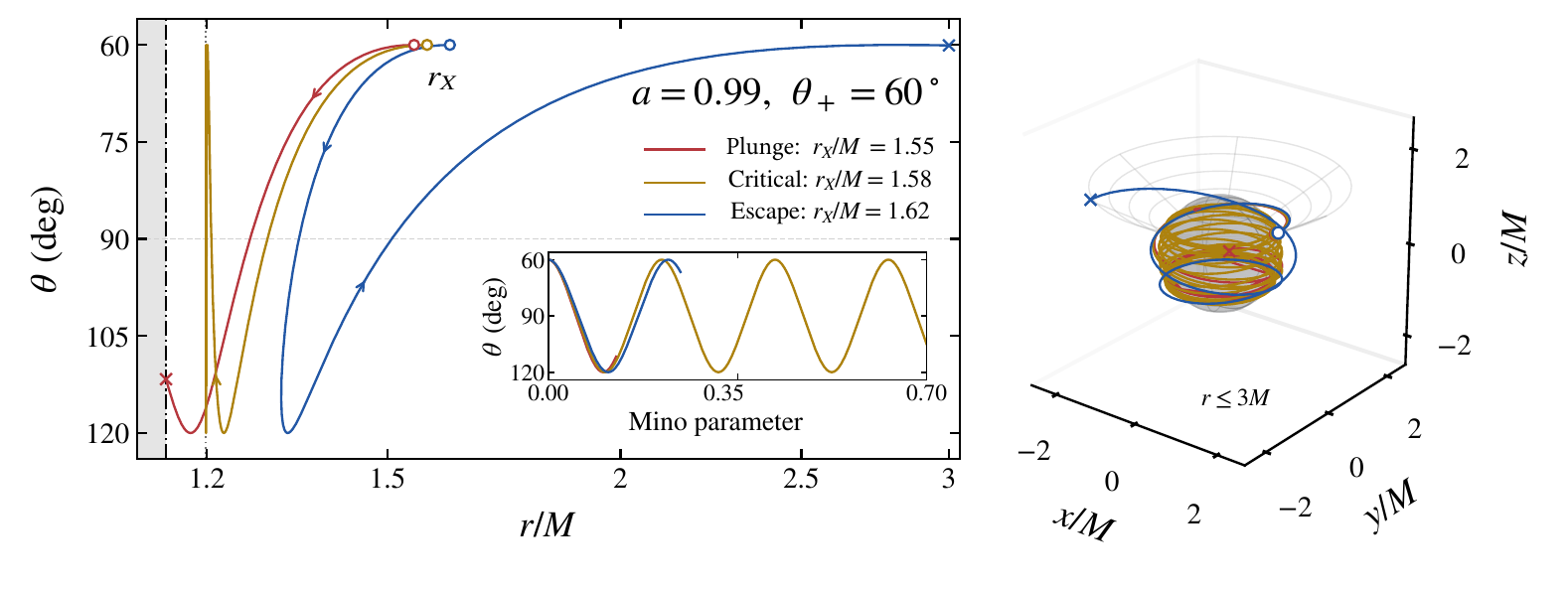}
\caption{Trajectories of positive-energy plasmoids ejected at reconnection radii $r_X/M = 1.55, 1.58, 1.62$, for $a = 0.99$ and $\theta_+ = 60^\circ$. The left panel shows the trajectories in the $(r,\theta)$ plane for $r \leq 3M$, with an inset highlighting the polar oscillations. The right panel shows the three-dimensional trajectories, illustrating the azimuthal winding around the black hole.}
	\label{fig_tra}
\end{figure}

The escape--capture separatrix consists of trajectories that asymptote to an unstable spherical photon orbit at $r=\tilde r$. Such an orbit satisfies $\tilde R(\tilde r)=0$ and $\partial_r\tilde R(\tilde r)=0$, where $\tilde R(r)\equiv R(r)|_{\lambda=\tilde\lambda,q=\tilde q}$. Its conserved quantities are
\begin{equation}
\tilde \lambda=\frac{\tilde r^2(\tilde r-3)+a^2(\tilde r+1)}{a(1-\tilde r)}\,,\quad
\tilde q=\frac{\tilde r^3[4a^2-\tilde r(\tilde r-3)^2]}{a^2(\tilde r-1)^2}\,.
\end{equation}
For the positive-energy branch considered here, the critical emission radius $r_{\rm cap}$ is therefore obtained by matching
$\lambda_+(r_{\rm cap},\theta_+,a)=\tilde\lambda(\tilde r,a)$ and
$q_+(r_{\rm cap},\theta_+,a)=\tilde q(\tilde r,a)$.
Only solutions with an accessible unstable orbit, admissible polar motion, and a future-directed initial radial branch approaching that orbit belong to the separatrix. Exactly critical trajectories asymptote to the spherical photon orbit and neither escape to infinity nor cross the horizon. Representative trajectories near the escape boundary are shown in Fig.~\ref{fig_tra}.

The matching conditions can also be expressed directly in terms of the upstream magnetic field. The kernel equation $F_{\mu\nu}N_+^\nu=0$ implies
$N_+^r\partial_r\psi+N_+^\theta\partial_\theta\psi=0$ and
$N_+^\phi-\Omega_F N_+^t=F_{r\theta}N_+^r(\partial_\theta\psi)^{-1}$.
For the split-monopole poloidal field considered here, $\partial_r\psi=0$ and $\partial_\theta\psi\neq0$, so the initial polar momentum vanishes. Using
\(F_{r\theta}=J\Sigma/(\Delta\sin\theta)\), where
\(\Sigma=r^2+a^2\cos^2\theta\), together with
the null-geodesic potentials, we obtain
\begin{equation}
	\label{eq:cap_condi}
	\begin{aligned}
&\tilde q-\tilde \lambda^2\cot^2\theta_+ + a^2\cos^2\theta_+ =0\,,\\
&\partial_\theta\psi\left[(g_{t\phi}+\Omega_Fg_{\phi\phi})+(g_{tt}+\Omega_Fg_{t\phi})\tilde \lambda\right]
+\operatorname{sgn}(N_+^r)\,J\sin\theta_+\sqrt{\tilde R(r_{\rm cap})}=0\,,
	\end{aligned}
\end{equation}
where all metric components and magnetic quantities are evaluated at $(r,\theta)=(r_{\rm cap},\theta_+)$. The first relation imposes vanishing initial polar momentum, whereas the second enforces the azimuthal kernel condition with the appropriate radial branch. Together, they determine candidate pairs $(r_{\rm cap},\tilde r)$, showing how the escape boundary depends jointly on the upstream magnetic configuration and the critical photon orbit. Each candidate must satisfy the accessibility and branch conditions stated above. For more general poloidal geometries, the nonzero initial polar momentum and its sign must instead be retained through the full kernel equation. At points where $\partial_\theta\psi=0$, or where the initial radial momentum vanishes, the uneliminated kernel and geodesic equations provide a regular formulation.

\end{document}